\documentclass[preprint,preprintnumbers,amsmath,amssymb]{revtex4}
\usepackage{dcolumn}% Align table columns on decimal point
\usepackage{bm}% bold math
 
\usepackage{amsmath}  
\usepackage{amsfonts}   
\usepackage{graphics}    
\usepackage{psfrag} 
\usepackage{graphicx} 
\usepackage{epsfig}    
\usepackage{rotating} 
 \usepackage[latin1]{inputenc} 
 \usepackage{color}
 \usepackage{rotating} 
\begin{document}

\preprint{}

\title{
Mobility promotes and jeopardizes biodiversity in rock-paper-scissors games}% 


\author{Tobias Reichenbach, Mauro Mobilia, and Erwin Frey}
\affiliation{Arnold Sommerfeld Center for Theoretical Physics (ASC) and
  Center for NanoScience (CeNS), Department of Physics,
  Ludwig-Maximilians-Universit\"at M\"unchen, Theresienstrasse 37,
  D-80333 M\"unchen, Germany}

%\email{tobias.reichenbach@physik.lmu.de}
%\affiliation{%     

\date{\today}% It is always \today, today,

%\pacs{}% PACS, the Physics and Astronomy

\maketitle  
			     
	\newpage  
{\bf
Biodiversity is essential to the viability of ecological systems. Species diversity in ecosystems is promoted by cyclic, non-hierarchical interactions among competing populations. Central features of such non-transitive relations are represented by the `rock-paper-scissors' game, where rock crushes scissors, scissors cut paper, and paper wraps rock. In combination with spatial dispersal of  static populations, this type of competition results in the stable coexistence of all species and the long-term maintenance of biodiversity$^{\ref{durrett-1997-185}-\ref{szabo-2006}}$. However, population mobility is a central feature of real ecosystems: animals migrate, bacteria run and tumble. Here, we observe a critical influence of mobility on species diversity. When mobility exceeds a certain value, biodiversity is jeopardized and lost. In contrast, below this critical threshold all subpopulations coexist and an entanglement of travelling spiral waves forms in the course of time. We establish that this phenomenon is robust,  it does not depend on the details of  cyclic competition or spatial environment.
These findings have important implications for maintenance and temporal development of ecological systems and are relevant for the formation and propagation of patterns in microbial  populations or excitable media. 
}

The remarkable biodiversity present in ecosystems confounds a na\" {i}ve interpretation of Darwinian evolution in which interacting species compete for limited resources until only the fitter species survives. 
As a striking example, consider that a 30-g sample of soil from a Norwegian forest is estimated to contain some 20,000 common bacterial species$^{\ref{dykhuizen-1998-73}}$.  Evolutionary game theory$^{\ref{Smith}-\ref{Nowak}}$, in which the success of one species relies on the behaviour of others, provides a useful framework to investigate co-development of populations theoretically. In this context, the rock-paper-scissors game has emerged as a paradigm to describe species diversity$^{\ref{durrett-1997-185}-\ref{szabo-2006},\ref{may-1975-29} - \ref{reichenbach-2006-74}}$. If three subpopulations interact in this non-hierarchical way, one expects intuitively that diversity may be preserved: Each species dominates another only to be outperformed by the remaining one in an endlessly spinning wheel of species-chasing-species.     

Communities of subpopulations exhibiting such dynamics have been identified in numerous ecosystems, ranging from coral reef invertebrates$^{\ref{jackson-1975-72}}$ to lizards in the inner Coast Range of California$^{\ref{sinervo-1996-340}}$. In particular, recent experimental studies using microbial laboratory cultures have been devoted to the influence of spatial structure on time development and coexistence of species$^{\ref{kerr-2002-418},\ref{kirkup-2004-428}}$.  Investigating three strains of colicinogenic \emph{E.coli} in different environments, it has been shown that cyclic dominance alone is not sufficient to preserve biodiversity. Only when the interactions between individuals are local (e.g. bacteria  arranged on a Petri dish), spatially separated domains dominated by one subpopulation  can form and lead to stable  coexistence$^{\ref{durrett-1997-185},\ref{kerr-2002-418}}$. 

In this Letter, we show that biodiversity is affected drastically by spatial migration of individuals, a ubiquitous feature of real ecosystems. Migration competes with local interactions such as reproduction and selection, thereby mediating species preservation and biodiversity. For low values of mobility, the temporal development is dominated by interactions among neighbouring individuals, resulting in the long-term maintenance of species diversity. In contrast, when species mobility is high, spatial homogeneity results and biodiversity is lost. Interestingly, a critical value of mobility sharply delineates these two scenarios. We obtain concise predictions for the fate of the ecological system as a function of species mobility, thereby gaining a comprehensive understanding of its biodiversity.

The influence of mobility on species coexistence was previously studied within the framework of coupled habitat patches (``island models'')$^{\ref{levin-1974-108}-\ref{king-2003-64}}$.  In particular, Levin  considered an idealized two-patch system and observed a critical mobility for stable coexistence$^{\ref{levin-1974-108}}$. Other models comprising many spatially arranged patches were shown to facilitate pattern formation$^{\ref{hassell-1991-353},\ref{blasius-1999-399}}$. 
As often in nature spatial degrees of freedom vary continuously (e.g. bacteria can visit the entire area of a Petri dish),  
we relax the simplifying assumption of habitat patches and consider continuous spatial distribution of individuals. Moreover, as an inherent feature of real ecosystems and in contrast to previous
deterministic investigations$^{\ref{levin-1974-108}-\ref{king-2003-64}}$, we explicitly take the
stochastic character of the interactions among the populations into account. Such interacting particle systems, where individuals are discrete and space is
treated explicitly, were already considered in ecological contexts$^{\ref{durrett-1997-185},\ref{durrett-1998-53},\ref{czaran-2002-99},\ref{szabo-2006},\ref{durrett-1994-46}}$. The behaviour of these
models often differs from what is inferred from deterministic reaction-diffusion
equations, or from  interconnected patches$^{\ref{durrett-1994-46}}$. In the case of cyclic
competition, such stochastic spatial systems have been shown to allow for stable coexistence of all species$^{\ref{durrett-1997-185},\ref{durrett-1998-53},\ref{czaran-2002-99},\ref{szabo-2006}}$ when individuals are static. Here we explore the novel features  emerging from individuals' mobility.

Consider mobile individuals of three subpopulations (referred to as $A$, $B$, and $C$), arranged on a spatial lattice, where they can only interact with nearest neighbours. For the possible interactions, we consider a version of the rock-paper-scissors game, namely
a stochastic spatial variant of the model introduced in 1975 by May and Leonard$^{\ref{may-1975-29}}$ (see Methods). Schematic illustrations of the model's dynamics are  provided in Fig.~1. 
The basic reactions comprise selection and reproduction processes, which occur at rates $\sigma$ and $\mu$, respectively.  Individuals' mobility stems from 
the possibility for two neighbouring individuals to swap their position (at rate $\epsilon$) and to move to an adjacent empty site.
Thereby, individuals randomly migrate on the lattice. We define the length of the square lattice as the size unit, and denote by $N$  the number of sites. Within this setting, and applying the theory of random walks$^{\ref{Redner}}$, the typical area explored by one mobile individual per unit time is proportional to $M=2\epsilon N^{-1}$, which we refer to as the \emph{mobility}.  The interplay of the latter with selection and reproduction processes sensitively determines whether species can coexist on the lattice or not, as discussed  below.

We performed  extensive computer simulations of the stochastic system (see Methods) and typical snapshots of the steady states are reported in Fig.~2. When the mobility of the individuals is low, we find that all species coexist and self-arrange by forming patterns 
of moving spirals. Increasing the mobility $M$, these structures grow in size, and disappear for large enough $M$. In the absence of spirals, the system adopts a uniform state where only one species is present, while the others have died out. 
Which species remains is subject to a random process, all species having equal chances to survive in our model.

Concise predictions on the stability of three-species coexistence are obtained by adapting the concept of \emph{extensivity} 
from statistical physics (see Supplementary Information).  Namely, we consider the typical waiting time $T$ until extinction occurs, and its dependence on the system size $N$. If $T(N)\sim N$, the stability of coexistence is marginal$^{\ref{reichenbach-2006-74}}$.  Conversely, longer (shorter) waiting times scaling with higher (lower) powers of $N$ indicate stable (unstable) coexistence.  These three  scenarios can be distinguished by computing the probability $P_\text{ext}$ that two species have gone extinct after a waiting time $t\sim N$. In Fig.~2, we report the dependence of $P_\text{ext}$ on the mobility $M$.
For illustration, we have considered equal reaction rates for selection and reproduction, and, without loss of generality, set the time-unit by fixing $\sigma=\mu=1$.
Increasing the system size $N$, a sharpened transition emerges at a critical value $M_c=(4.5\pm 0.5)\times 10^{-4}$ for the fraction of the entire lattice area explored by an individual in one time-unit.
Below $M_c$, the extinction probability $P_\text{ext}$ tends to zero as the system size increases,  and coexistence is stable (implying super-persistent transients$^{\ref{hastings-2004-19}}$, see Supplementary Information). On the other hand, above the critical mobility, the extinction probability approaches one for large system size, and coexistence is unstable.  As a central result of this Letter, we have identified a \emph{mobility threshold} for biodiversity: 
    
\emph{There exists a critical value $M_c$ such that a low mobility $M<M_c$ guarantees coexistence of all three species, while $M>M_c$ induces extinction of two of them, leaving a uniform state with only one species.} 

To give a biological illustration of this statement, let us consider colicinogenic strains of \emph{E.coli}  growing on a Petri dish$^{\ref{kerr-2002-418}}$. In this setting, 10 bacterial generations have been observed in 24 hours, yielding selection and reproduction rates 
of about 10 per day. As the typical size of a Petri dish is roughly $10~\text{cm}$, we have evaluated the critical mobility to be about $5\times 10^2~\mu\text{m}^2/\text{s}$. 
Comparing that estimate to the mobility of {\it E.coli}, we find that it can,  by swimming and tumbling in super soft agar, explore areas of more than $10^3~\mu\text{m}^2$  per second$^{\ref{Berg}}$. This value can be considerably lowered by increasing the agar concentration.

When the mobility is low, i.e. $M<M_c$,  the interacting subpopulations exhibit fascinating patterns,
as illustrated by the snapshots of Fig.~2. The emerging reactive states are formed by an entanglement of spiral waves,  characterising the competition among the species which endlessly hunt each other, as illustrated in movies 1 and 2 (see Supplementary Information).  Formation of this type of patterns has been observed in microbial populations, such as myxobacteria aggregation$^{\ref{igoshin-2004-101}}$ or multicellular {\it Dictyostelium} mounds$^{\ref{siegert-1995-937}}$, as well as in cell signaling and control$^{\ref{thul-2004-93}}$. Remarkably, a mathematical description  and techniques borrowed from the theory of stochastic processes$^{\ref{Gardiner}}$ allow to obtain these complex structures by means of stochastic partial differential equations (PDE), see Fig.~3 and Methods. Furthermore, recasting the dynamics in the form of a complex Ginzburg-Landau equation$^{\ref{Wiggins},\ref{aranson-2002-74}}$ 
allows to obtain analytical expressions for the  spirals' wavelength $\lambda$ and frequency (see Supplementary Information). 
These results, up to a constant prefactor, agree with those of numerical computations; see our forthcoming publication (T.R., M.M. and E.F., in preparation).

As shown in  Fig.~2, the spirals' wavelength $\lambda$ rises with the individuals' mobility.
Our analysis reveals that the wavelength is proportional to $\sqrt{M}$ (see Supplementary Information). This  relation  holds up to the mobility  $M_c$, where a \emph{critical wavelength} $\lambda_c$ is reached. For mobilities above the threshold $M_c$,  the spirals' wavelength $\lambda$ exceeds the critical value $\lambda_c$ and the patterns outgrow the system size causing 
the loss of biodiversity, see Fig.~2.
We have found  $\lambda_c$ to be universal, i.e. independent on the selection and reproduction rates.
This is not the case for $M_c$, whose value varies with these parameters (see Supplementary Information). 
Using lattice simulations, stochastic PDE and the properties of the complex Ginzburg-Landau equation, we have derived the 
dependence of the critical mobility $M_c(\mu)$ on the reproduction rate $\mu$ (where the time-unit is set by keeping $\sigma=1$). It enables to analytically predict, for all values of parameters, whether biodiversity is maintained or lost. We have summarized these results in a phase diagram, reported in Fig.~4. One identifies a uniform phase, in which two species go extinct (when $M>M_c(\mu)$), and a biodiverse phase (when $M<M_c(\mu)$) with coexistence of all species and propagation of spiral waves.

The generic ingredients for the above scenario to hold are the mobility of the individuals
and a cyclic dynamics exhibiting an unstable reactive fixed point. 
The underlying mathematical description of this class of dynamical systems is in terms of complex Ginzburg-Landau equations. Their universality classes reveal the robustness of the phenomena which we have reported above, i.e. the existence 
of a critical mobility and the emergence of spiral waves; they are  not restricted to  specific details of the model.

Our study has direct implications for experimental research on biodiversity and pattern formation. As an example, one can envisage an experiment extending the study$^{\ref{kerr-2002-418}}$ on  colicinogenic \emph{E.coli}. Allowing the bacteria to migrate in soft agar on a Petri dish should, for low mobilities, result in stable coexistence promoted by the formation of spiral patterns. Increasing the mobility (e.g. on super soft agar), the patterns should grow in size and finally outgrow the system at some critical value, corresponding to the threshold $M_c$ discussed in this Letter. For even higher values of the mobility, biodiversity should be lost after a short transient time and only one species should cover the entire Petri dish. We think that both mobility's regimes, corresponding to the biodiverse and uniform phases, should be experimentally accessible.

We have shown how concepts from game theory combined with methods used to study pattern formation reveal the subtle influence
of mobility on the temporal development of coexisting species. Many more questions and applications regarding the seminal interplay between these different fields lie ahead. As an example, concerning the evolution of cooperation, it has been shown that cyclic dominance can occur in social dilemmas$^{\ref{Nowak},\ref{hauert-2002-296}}$, which suggests implications of our results to behavioural sciences.

\section*{Methods}

To model cyclic dominance, we use a stochastic lattice version (following work by Durrett and Levin) of a model proposed by May and Leonard in 1975. As main characteristics, in the absence of spatial structure, their equations possess a deterministically {\it unstable} fixed point associated to coexistence of all three species: In the course of time, the system spirals (in the phase space) away from coexistence and moves in turn from a state with nearly only $A$'s to another one with nearly only $B$'s, and then to a state with almost only $C$'s.

In our stochastic lattice simulations, we have arranged the three subpopulations on a two-dimensional square lattice with periodic boundary conditions. Every lattice site is occupied by an individual of species $A,B,$ or $C$, or left empty. At each simulation step, a random individual is chosen to interact with one of its four nearest neighbours, which is also randomly determined. Whether selection, reproduction or mobility occurs, as well as the corresponding waiting time, is computed according to the reaction rates using an efficient algorithm due to Gillespie$^{\ref{gillespie-1976-22}}$. We set one generation, i.e. when every 
individual has reacted on average once, as unit of time. To compute the extinction probability, we have used different system sizes, from $20\times 20$ to $200\times 200$ lattice sites, and sampled between 500 and 2000 realizations. The snapshots shown in Fig.~2 result from system sizes of up to $1000\times 1000$ sites.

Our stochastic PDE consist of a mobility term, nonlinear terms describing the deterministic temporal development of the nonspatial model (May-Leonard equations), and (multiplicative) white noise, see Supplementary Information. 
We have solved the resulting equations with the help of open software from the XMDS project$^{\ref{xmds}}$, using the semi-implicit method in the interaction picture (SIIP) as an algorithm,  spatial meshes of $200\times 200$ to $500\times 500$ points, and $10,000$ points in the time direction.

\subsection*{Supplementary Information} is available online.

\subsection*{Acknowledgements}
We thank M. Bathe and M. Leisner for inspiring discussions and helpful comments on the manuscript.
Financial support of the German Excellence Initiative via the program ``Nanosystems Initiative Munich (NIM)'' is gratefully acknowledged. M. M.   is grateful to the Alexander von Humboldt Foundation
for support through a fellowship. 

\subsection*{Competing interest statement}
 
The authors declare they have no competing financial interest. Correspondence should be addressed to E. F. (frey@lmu.de).

\newpage

{\bf \large References}

\begin{enumerate}

\item
\label{durrett-1997-185}
Durrett, R. \&  Levin, S.
\newblock Allelopathy in spatially distributed populations.
\newblock {\em J. Theor. Biol.} {\bf 185}, 165-171  (1997).

\item
\label{durrett-1998-53}
Durrett, R. \&  Levin, S.
\newblock Spatial aspects of interspecific competition.
\newblock {\em Theor. Pop. Biol.}  {\bf 53}, 30-43  (1998).

\item
\label{kerr-2002-418}
Kerr, B., Riley, M. A., Feldman, M. W. \& Bohannan, B. J. M.
\newblock Local dispersal promotes biodiversity in a real-life game of
  rock-paper-scissors.
\newblock {\em Nature}  {\bf 418}, 171-174 (2002).

\item
\label{czaran-2002-99}
Cz\'ar\'an, T. L., Hoekstra, R. F. \& Pagie, L.
\newblock Chemical warfare between microbes promotes biodiversity.
\newblock {\em PNAS} {\bf 99},  786-790 (2002).

\item
\label{szabo-2006}
Szab\'o , G. \&  F\'ath, G.
\newblock Evolutionary games on graphs.
\newblock {\em Phys. Rep.} {\bf 446}, 97-216 (2007).

\item \label{dykhuizen-1998-73}
Dykhuizen, D.~E.
\newblock Santa rosalia revisited: Why are there so many species of bacteria?
\newblock {\em Antonie Leeuwenhoek} {\bf 73}, 25-33 (1998).

\item
\label{Smith}
Smith, J.~M.
\newblock {\em Evolution and the Theory of Games}
\newblock (Cambridge Univ. Press, Cambridge, 1982).

\item
\label{Hofbauer}
Hofbauer, J. \& Sigmund, K.
\newblock {\em Evolutionary Games and Population Dynamics}
\newblock (Cambridge Univ. Press, Cambdrige 1998).

\item
\label{Nowak}
Nowak, M.~A.
\newblock {\em Evolutionary Dynamics}
\newblock (Belknap Press, Cambridge MA, 2006).

\item
\label{may-1975-29} 
May, R. M. \& Leonard, W. J. 
\newblock Nonlinear aspects of competition between species.
\newblock {\em SIAM J. Appl. Math}  {\bf 29}, 243-253  (1975).

\item
\label{johnson-2002-269}
Johnson, C. R. \&  Seinen, I.
\newblock Selection for restraint in competitive ability in spatial competition
  systems.
\newblock {\em Proc. R. Soc. Lond. B}  {\bf 269}, 655-663 (2002).

\item
\label{reichenbach-2006-74}
Reichenbach, T., Mobilia, M. \& Frey, E.
\newblock Coexistence versus extinction in the stochastic cyclic Lotka-Volterra
  model.
\newblock {\em Phys. Rev. E}  {\bf 74}, 051907  (2006).

\item
\label{jackson-1975-72}
Jackson, J. B. C. \& Buss, L.
\newblock Allelopathy and spatial competition among coral reef invertebrates.
\newblock {\em PNAS} {\bf 72}, 5160-5163  (1975).

\item
\label{sinervo-1996-340}
Sinervo, B. \& Lively, C. M.
\newblock The rock-scissors-paper game and the evolution of alternative male
  strategies.
\newblock {\em Nature}  {\bf 380}, 240-243  (1996).

\item
\label{kirkup-2004-428}
Kirkup, B. C. \&  Riley, M. A.
\newblock Antibiotic-mediated antagonism leads to a bacterial game of
  rock-paper-scissors \emph{in vivo}.
\newblock {\em Nature}  {\bf 428}, 412-414  (2004).

%\item
%\label{kerr-2006-442}
%Kerr, B., Neuhauser, C., Bohannan, B. J. M. \& Dean, A. M.
%\newblock Local migration promotes competitive restraint in a host-pathogen
%  `tragedy of the commons'.
%\newblock {\em Nature}  {\bf 442}, 75-78  (2006).

\item
\label{levin-1974-108}
Levin, S. A.
\newblock Dispersion and population interactions.
\newblock {\em Am. Nat.}  {\bf 108}, 207-228  (1974).

\item
\label{hassell-1991-353}
Hassell, P. M., Comins, H. N., \& May, R. M.
\newblock Spatial structure and chaos in insect population dynamics.
\newblock {\em Nature}  {\bf 353}, 255-258 (1991).

\item
\label{blasius-1999-399}
Blasius, B., Huppert, A., \&  Stone, L.
\newblock Complex dynamics and phase synchronization in spatially extended
  ecological systems.
\newblock {\em Nature}  {\bf 399}, 354-359  (1999).

\item
\label{king-2003-64}
King, A. A. \& Hastings, A.
\newblock Spatial mechanism for coexistence of species sharing a common natural
  enemy.
\newblock {\em Theor. Pop. Biol.}  {\bf 64}, 431-438 (2003).

\item
\label{durrett-1994-46}
Durrett, R. \&  Levin, S.
\newblock The importance of being discrete (and spatial).
\newblock {\em Theor. Pop. Biol.}  {\bf 46}, 363-394 (1994).

\item
\label{Redner}
Redner, S.
\newblock {\em A guide to first-passage processes}
\newblock (Cambridge Univ. Press, Cambridge, 2001).

%\item
%\label{frey-2005-14}
%Frey, E. \& Kroy, K.
%\newblock Brownian motion: a paradigm of soft matter and biological physics.
%\newblock {\em Ann. d. Physik}  {\bf 14}, 20-50  (2005).

\item
\label{hastings-2004-19}
Hastings, A.
\newblock Transients: the key to long-term ecological understanding?
\newblock {\em Trends Ecol. Evol.}  {\bf 19}, 39-45  (2004).

\item
\label{Berg}
Berg, H. C.
\newblock {\em \emph{E. coli} in Motion}
\newblock (Springer, New York, 2003).

\item
\label{igoshin-2004-101}
Igoshin, O. A., Welch, R., Kaiser, D., \& Oster, G.
\newblock Waves and aggregation patterns in myxobacteria.
\newblock {\em PNAS}  {\bf 101}, 4256-4261 (2004).

\item 
\label{siegert-1995-937}
Siegert, F. \& Weijer, C. J.  
\newblock Spiral and concentric waves organize multicellular {\it
  Dictyostelium} mounds.
\newblock {\em Curr. Biol.}  {\bf 5}, 937-943  (1995).

\item
\label{thul-2004-93}
Thul, R. \&  Falcke, M.
\newblock Stability of membrane bound reactions.
\newblock {\em Phys. Rev. Lett.}  {\bf 93}, 188103  (2004).

\item
\label{Gardiner}
Gardiner, C. W.
\newblock {\em Handbook of Stochastic Methods}
\newblock (Springer, Berlin, 1983).

\item
\label{Wiggins}
Wiggins, S.
\newblock {\em Introduction to Applied Nonlinear Dynamical Systems and Chaos}
\newblock (Springer, New York, 1990).

%\item
%\label{cross-1993-65}
%Cross, M. C. \&  Hohenberg, P. C..
%\newblock Pattern formation outside of equilibrium.
%\newblock {\em Rev. Mod. Phys.} {\bf 65}, 851-1112  (1993).

\item
\label{aranson-2002-74}
Aranson, I. S. \& Kramer, L.
\newblock The world of the complex Ginzburg-Landau equation.
\newblock {\em Rev. Mod. Phys.}  {\bf 74}, 99-143  (2002).

\item
\label{hauert-2002-296}
Hauert, C., de~Monte, S., Hofbauer, J. \&  Sigmund, K..
\newblock Volunteering as red queen mechanism for cooperation in public goods
  games.
\newblock {\em Science}  {\bf 296}, 1129-1132  (2002).

%\item
%\label{nowak-2005-427}
%Nowak, M. A. \& Sigmund, K.
%\newblock Evolution of indirect reciprocity.
%\newblock {\em Nature} {\bf 427}, 1291-1298 (2005).

%\item
%\label{grenfell-2001-414}
%Grenfell, B. T., Bjornstad, O. N. \& Kappey, J.
%\newblock Travelling waves and spatial hierarchies in measles epidemics.
%\newblock {\em Nature} {\bf 414}, 716-723  (2001).

%\item
%\label{cummings-2004-427}
%Cummings, D. A. T. et a.
%\newblock Travelling waves in the occurrence of dengue haemorrhagic fever in
%  thailand.
%\newblock {\em Nature}  {\bf 427}, 344-347  (2004).

\item
\label{gillespie-1976-22}
Gillespie, D. T.
\newblock A general method for numerically simulating the stochastic time evolution of coupled chemical reactions.
\newblock {\em J. Comp. Phys.} {\bf 22}, 403-434  (1976).

\item
\label{xmds}
\newblock{URL http://www.xmds.org}

%\item
%\label{collecutt-2001-142}
%Collecutt, G. R. \&  Drummond, P. D.
%\newblock xmds: extensible multi-dimensional simulator.
%\newblock {\em Comput. Phys. Commun.}  {\bf 142}, 219-223  (2001).

% new references... 

\end{enumerate}

%\bibliographystyle{unsrt}

% \bibliography{lit_rps}% Produces the bibliography via BibTeX.

%%%%%%%%%%%%%%%%%%%%%%%%%%%%%%%%%%%%%%%%%%%%%%%%%%%%%%%%%%%%%%%%%%%%%%%%%%%%%%%%%%%%%%%% 
% Figures
%%%%%%%%%%%%%%%%%%%%%%%%%%%%%%%%%%%%%%%%%%%%%%%%%%%%%%%%%%%%%%%%%%%%%%%%%%%%%%%%%%%%%%%%

\newpage

{\bf Figure captions}

\vspace{1cm}
{\small
{\bf Figure 1}\\
The rules of the stochastic model. Individuals of three competing species $A$ (red), $B$ (blue), and $C$ (yellow) occupy the sites of a lattice. {\bf\sffamily a}, They interact with their nearest neighbours through selection (i) or reproduction (ii). Both reactions occur as Poisson processes at rate $\sigma$ and $\mu$, respectively. Selection reflects cyclic dominance: $A$ can kill $B$, yielding an empty site (black) there. In the same way, $B$ invades $C$, and $C$ in turn outcompetes $A$. Reproduction of individuals is only allowed on empty neighbouring sites, mimicking a finite carrying capacity of the system.  We also endow individuals with mobility: at rate $\epsilon$, they are able to swap position with a neighbouring individual or hop on an empty neighbouring site (iii). {\bf\sffamily b}, An example of the processes (i)-(iii), taking place on a $3\times3$ square lattice. \label{reactions}

\vspace{1cm}
{\bf Figure 2}\\
The critical mobility $M_c$. Mobility below the value $M_c$ induces biodiversity; while it is lost above that threshold. {\bf\sffamily a}, We show snapshots obtained from lattice simulations of typical states of the system after long temporal development (i.e. at time $t\sim N$) and for different values of $M$  
(each color, blue, yellow and red, represents one of the species and black dots indicate empty spots). Increasing $M$  (from left to right), the spiral structures grow, and outgrow the system size at the critical mobility $M_c$: then, coexistence of all three species is lost and uniform populations remain (right). {\bf\sffamily b}, Quantitatively, we have considered the extinction probability $P_\text{ext}$ that, starting with randomly distributed individuals on a square lattice, the system has reached an absorbing state after a waiting time $t=N$. We compute $P_\text{ext}$ as function of the mobility $M$ (and $\sigma=\mu=1$), and show results for different system sizes: $N=20\times 20$ (green), $N=30\times 30$ (red), $N=40\times 40$  (purple), $N=100\times 100$ (blue), and $N=200\times 200$ (black)
 . As the system size increases, the  transition from stable coexistence ($P_\text{ext}=0$) to extinction ($P_\text{ext}=1$) sharpens  at a critical mobility  $M_c\approx(4.5\pm0.5)\times 10^{-4}$. \label{critical_D}

\newpage
{\bf Figure 3}\\
Spiralling patterns. {\bf\sffamily a}, Typical spiral (schematic). It rotates around the origin (white dot) at a frequency $\omega$ and possesses a wavelength $\lambda$. {\bf\sffamily b}, In our lattice simulations, when the mobility of individuals lies below the critical value, all three species coexist, forming  mosaics of entangled, rotating spirals (each color represents one of the species and black dots indicate empty spots). {\bf\sffamily c}, We have found that the system's development can aptly be described by stochastic partial differential equations. In the case of lattice simulations
and stochastic partial differential equations, internal noise acts as a source of local inhomogeneities and ensures the robustness of the dynamical behaviour: the spatio-temporal patterns are independent of the initial conditions. {\bf\sffamily d}, Ignoring the effects of noise, one is left with deterministic partial differential equations which also give rise to spiralling structures. The latter share the same wavelength and frequency with those of the stochastic description but, in the absence of fluctuations, their overall size and number depend on the initial conditions and can deviate significantly from their stochastic counterparts. In {\bf\sffamily b} and {\bf\sffamily c}, the system is initially in a homogeneous state, while
{\bf d} has been generated by considering an initial local perturbation. Parameters are $\sigma=\mu=1$ and $M=1\times 10^{-5}$.
\label{compare_snapshots}

\vspace{1cm} 
{\bf Figure 4}\\
Phase diagram. The critical mobility  $M_c$ as a function of the reproduction rate $\mu$ yields a phase diagram with a phase where biodiversity is maintained as well as a uniform one  where two species go extinct.  Time unit is set by $\sigma=1$. On the one hand, we have computed $M_c$ from lattice simulations, using different system sizes. The results are shown as blue crosses.  On the other hand, we have calculated $M_c$ using the approach of stochastic PDE (black dots, black lines are a guide to the eye) as well as analytically via the complex Ginzburg-Landau equation (red line).  Varying the reproduction rate, two different regimes emerge. If $\mu$ is much smaller than the selection rate, i.e. $\mu\ll\sigma$, reproduction is the dominant limiter of the temporal development. In this case, there
is a linear relation with the critical mobility, i.e. $M_c\sim\mu$, as follows from dimensional analysis. 
In the opposite case, if reproduction occurs much faster than selection ($\mu\gg\sigma$), the latter limits the dynamics and $M_c$ depends linearly on $\sigma$, i.e. $M_c\sim\sigma$. Here, as $\sigma=1$ is kept fixed (time-scale unit), this behaviour reflects in the fact that $M_c$ approaches a constant value for $\mu\gg\sigma$. \label{beta_D}

}

\newpage

{\bf Figures}

\vspace{1cm}

Figure 1

\vspace{2cm}

\begin{figure}[h]  
\begin{center} 
\psfrag{birth}{\hspace{-1.5cm}\sffamily(ii) Reproduction ($\mu$)}
\psfrag{kill}{\hspace{-1.8cm}\parbox{2cm}{\begin{center}\hspace{0.5cm}(i)\\\vspace{-0.2cm}\sffamily Selection~($\sigma$)\end{center}}}
\psfrag{exchange}{\hspace{-1.5cm}\sffamily(iii) Exchange ($\epsilon$)}
\begin{tabular}{lccc}
{\bf\sffamily a}\hspace{0.5cm} &\sffamily (i) Selection, rate $\sigma$: &$\qquad$& \sffamily (ii) Reproduction, rate $\mu$:\\ 
\vspace{-0.4cm}\\
&
\includegraphics[scale=0.23]{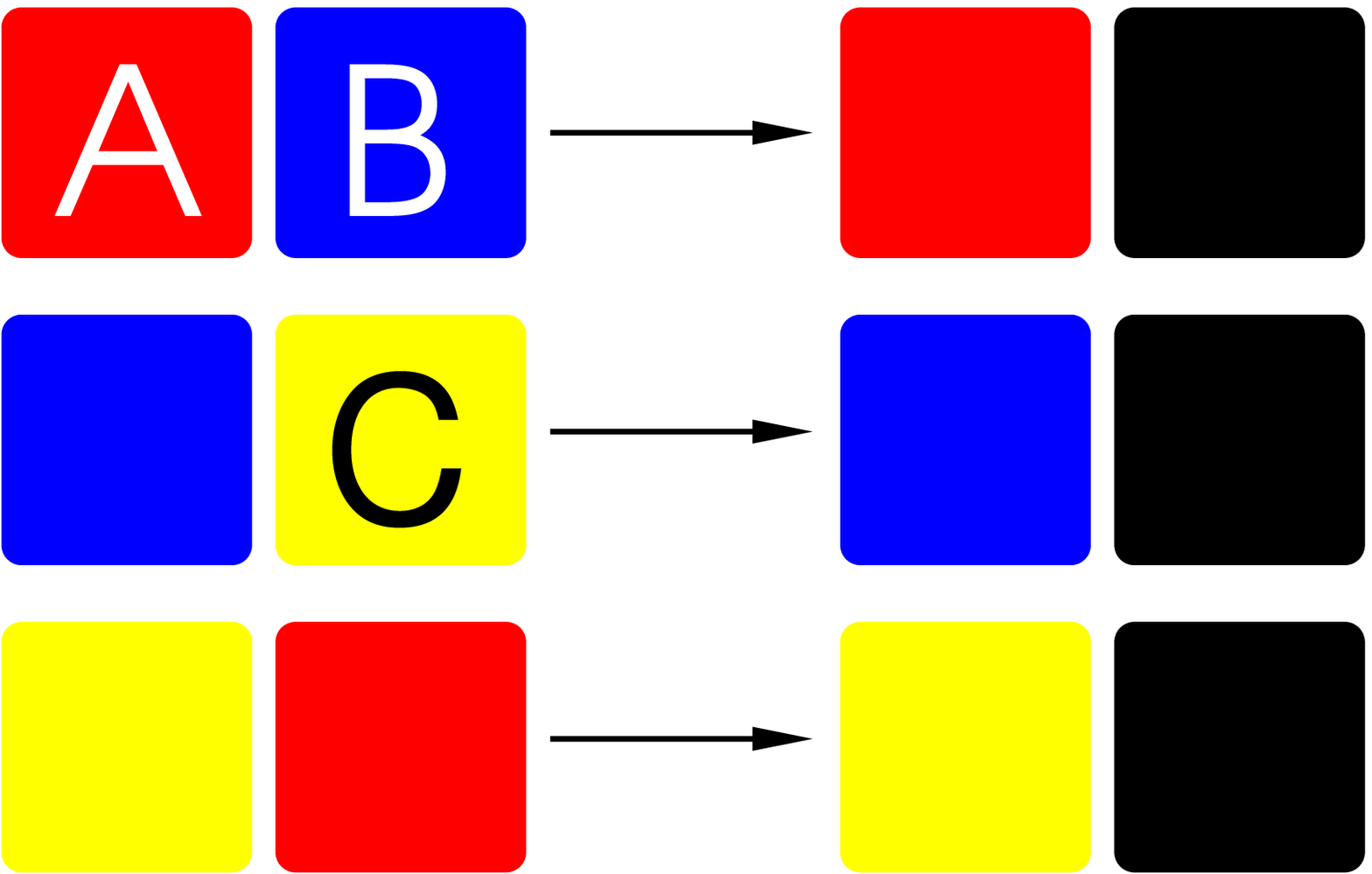}
&&
\includegraphics[scale=0.23]{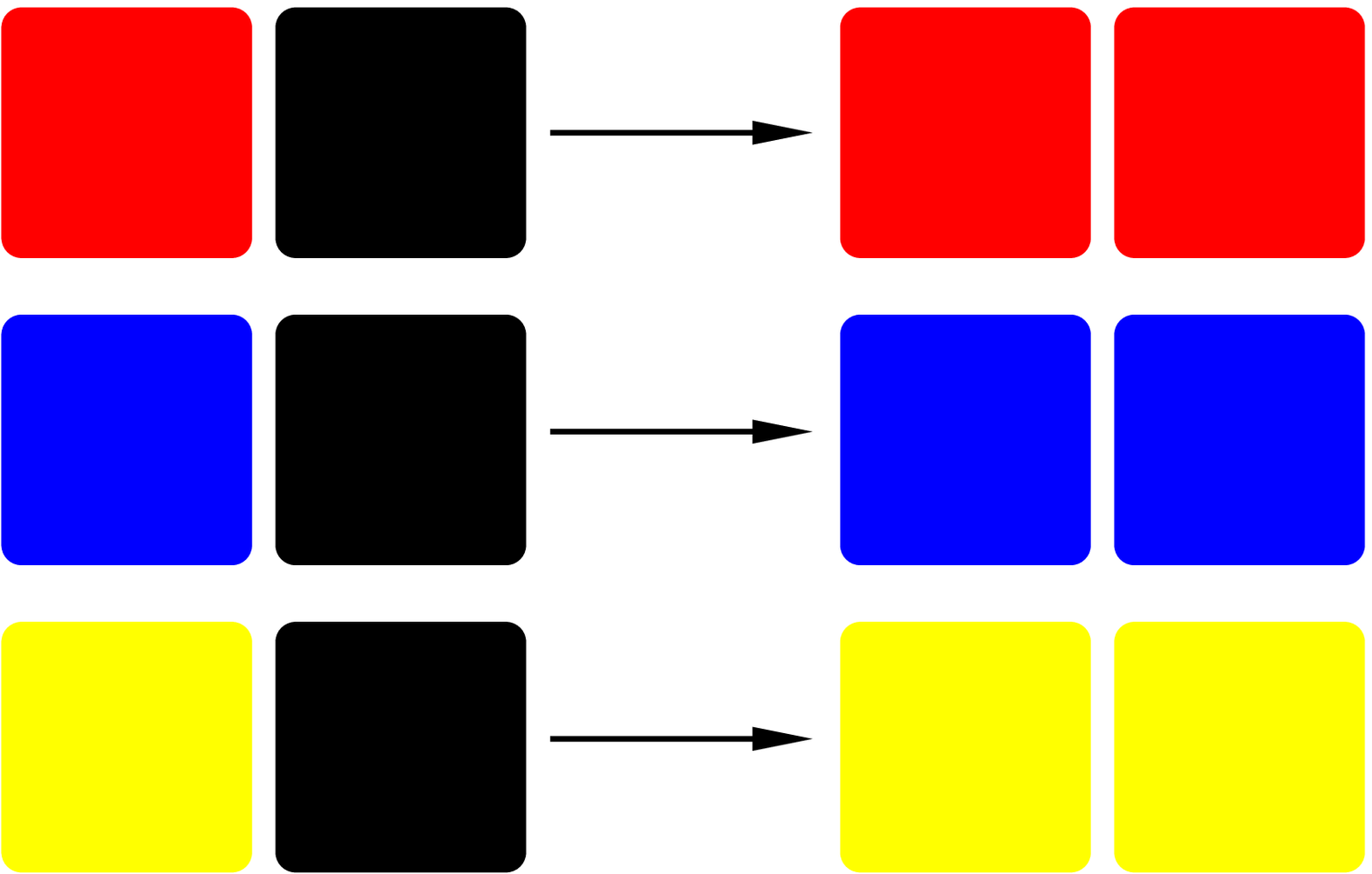}
\end{tabular}\\
\vspace{0.5cm}
\begin{tabular}{ccc}
\bf \hspace{-4cm}{\bf\sffamily b} &&\\ 
\parbox{5cm}{\vspace{0.2cm}
\includegraphics[scale=0.3]{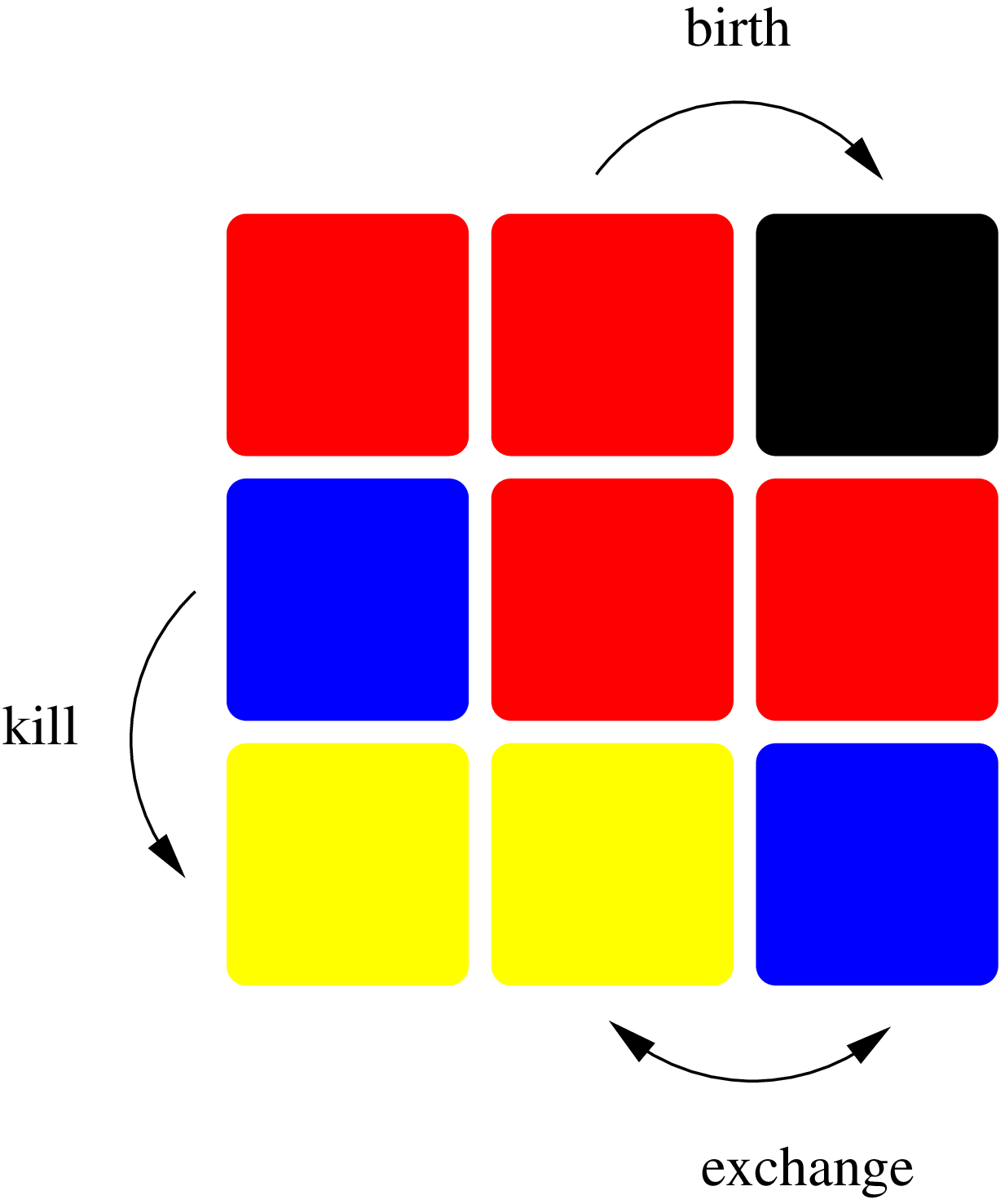}}
&\hspace{-0.4cm}\includegraphics[scale=0.25]{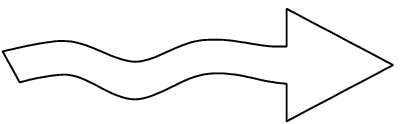}&\hspace{-0.7cm}
\parbox{5cm}{\vspace{0.2cm}
\hspace{-0.4cm}\includegraphics[scale=0.3]{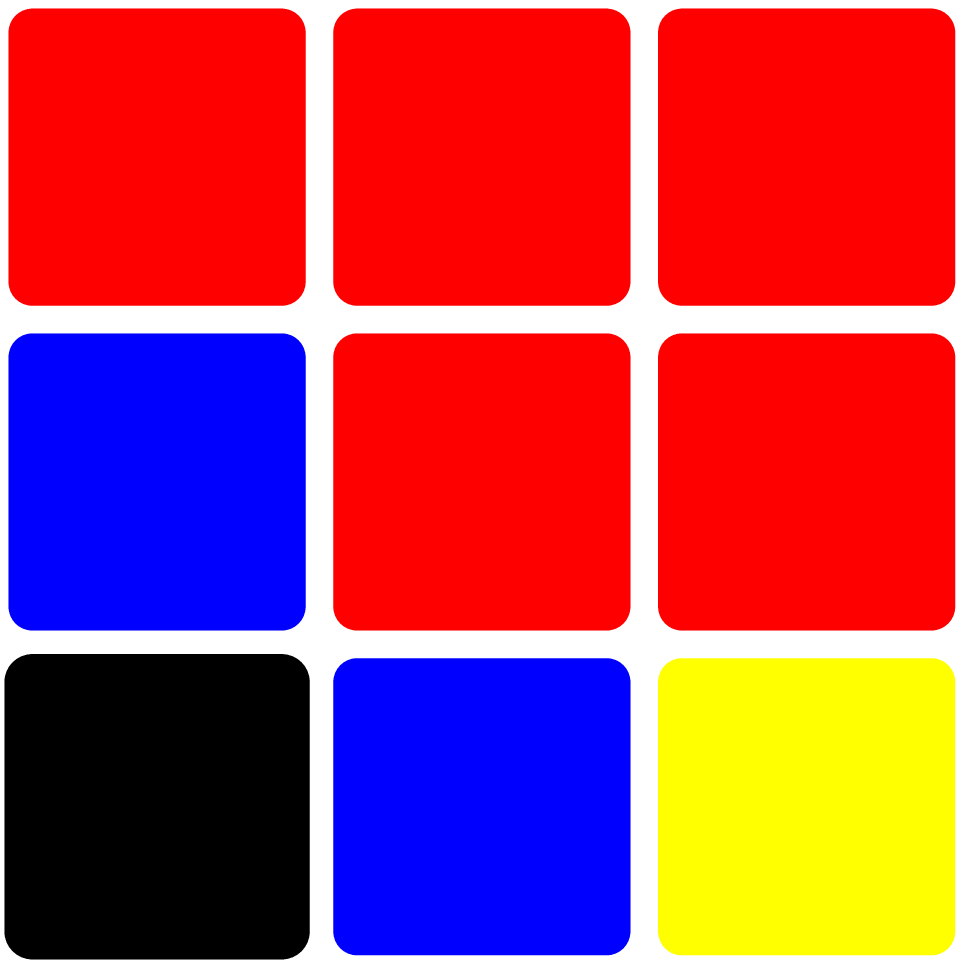}}
\end{tabular}
\end{center}
\end{figure}

\newpage

  Figure 2

\vspace{2cm}
   
\begin{figure}[h]
\begin{center}
\psfrag{a}{\bf\sffamily a}
\psfrag{b}{\bf\sffamily b}
\psfrag{Di}{$M$}
\psfrag{Dc}{$M_C$} 
\psfrag{3e-6}{\hspace{-0.5cm}$3\times 10^{-6}$}
\psfrag{3e-5}{\hspace{-0.5cm}$3\times 10^{-5}$}
\psfrag{3e-4}{\hspace{-0.5cm}$3\times 10^{-4}$}
\psfrag{-5}{\hspace{-0.2cm}\footnotesize $1\times 10^{-5}$}
\psfrag{-4}{\hspace{-0.2cm}\footnotesize $ 1\times 10^{-4}$}
\psfrag{-3}{\hspace{-0.2cm}\footnotesize $ 1\times 10^{-3}$}
\psfrag{0}{\hspace{-0.0cm}\footnotesize$0$}
\psfrag{2}{\hspace{-0.3cm}\footnotesize$0.2$}
\psfrag{4}{\hspace{-0.3cm}\footnotesize$0.4$}
\psfrag{6}{\hspace{-0.3cm}\footnotesize$0.6$}
\psfrag{8}{\hspace{-0.3cm}\footnotesize$0.8$}
\psfrag{1}{\hspace{-0.0cm}\footnotesize$1$ }
\psfrag{unstable}{ \hspace{-0.2cm}\sffamily Uniformity}
\psfrag{stable}{\sffamily Biodiversity}
\psfrag{Pext}{\begin{rotate}{90}\hspace{-3.5cm}\sffamily Extinction probability,~ $P_\text{ext}$\end{rotate}}
\psfrag{D}{\sffamily Mobility,~$M$}
\psfrag{Dc}{$M_c$} 
\includegraphics[scale=0.3]{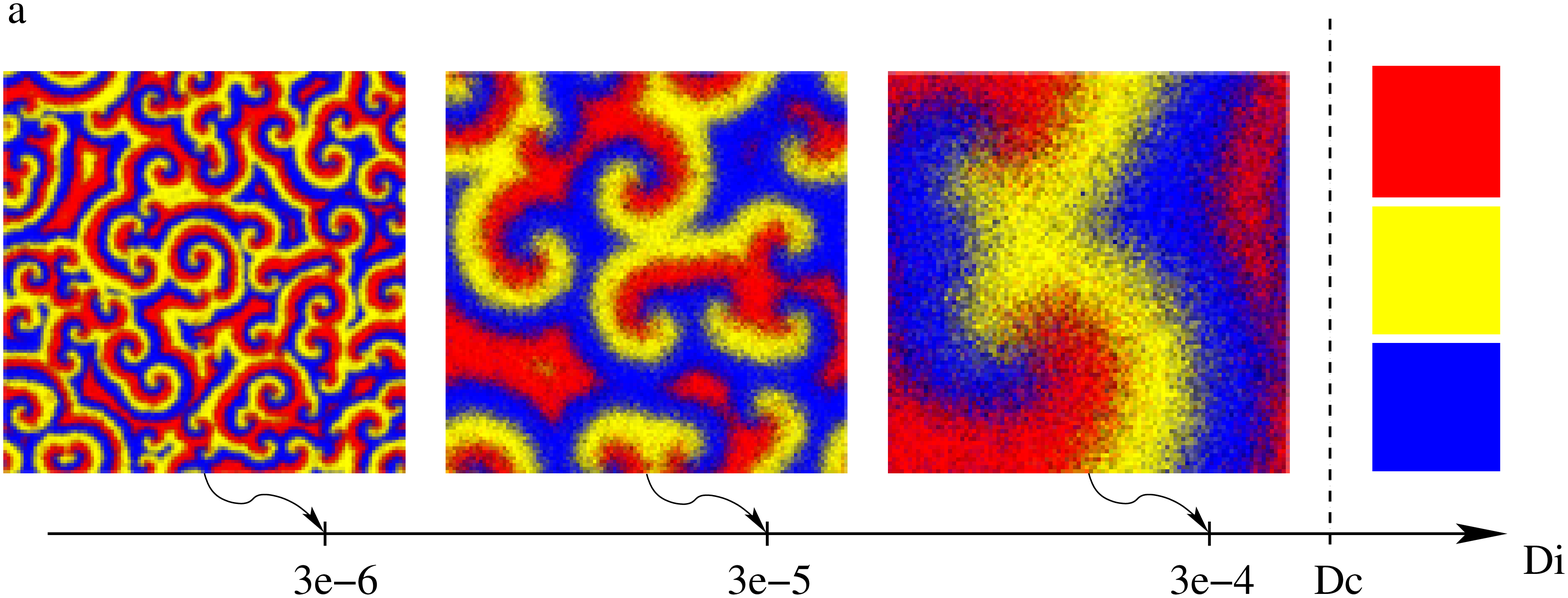}\\
\vspace{1cm}
\includegraphics[scale=0.7]{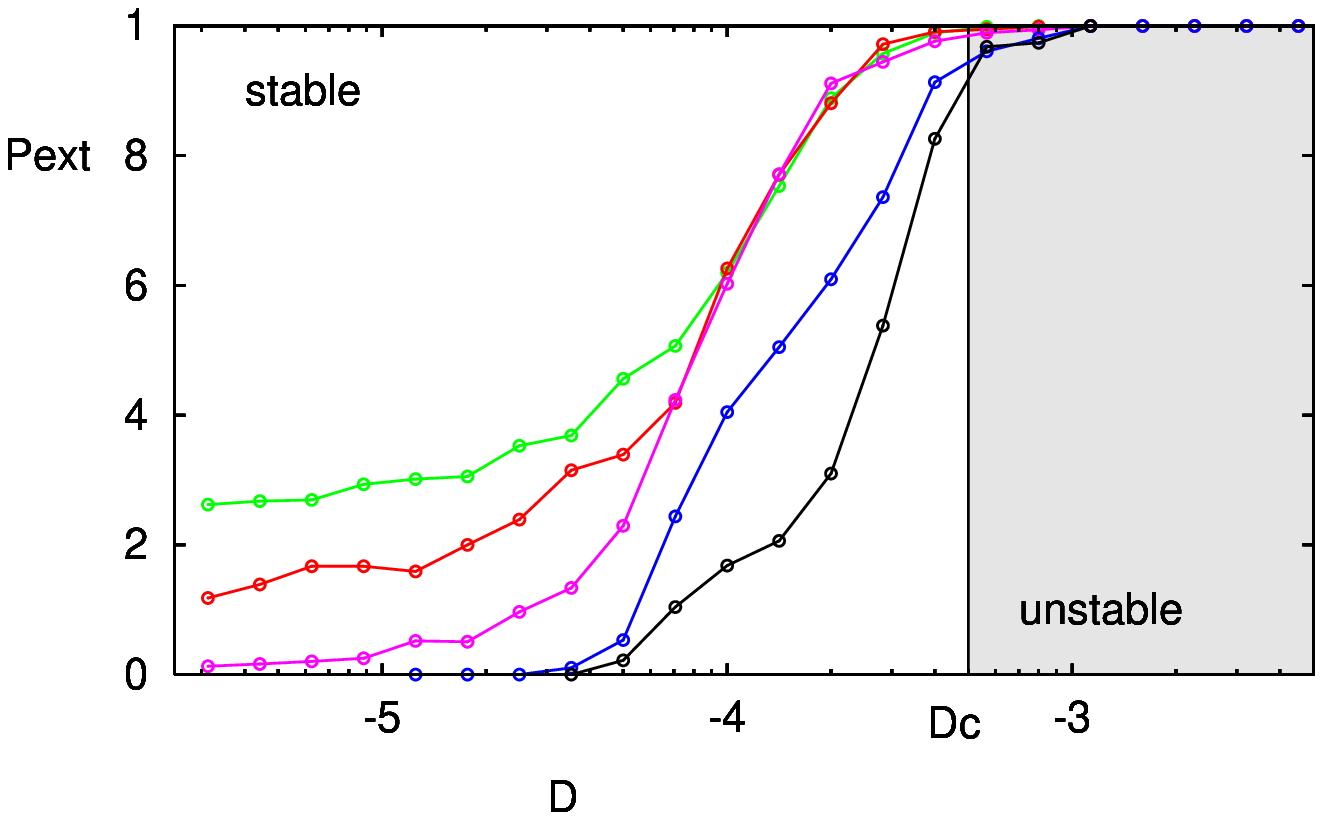}
\end{center}                 
\end{figure}

\newpage

   Figure 3

\vspace{2cm}
  
\begin{figure}[h]  
\psfrag{l}{\textcolor{white}{\large $\lambda$}}
\psfrag{o}{\textcolor{white}{\large $\omega$}}
\begin{center}   
\begin{tabular}{cccc}
{\bf \sffamily a}\sffamily~Typical spiral &{\bf\sffamily b}\sffamily~Lattice simulations & {\bf\sffamily c}\sffamily~Stochastic PDE & {\bf \sffamily d}~\sffamily Deterministic PDE\\
\includegraphics[scale=0.236]{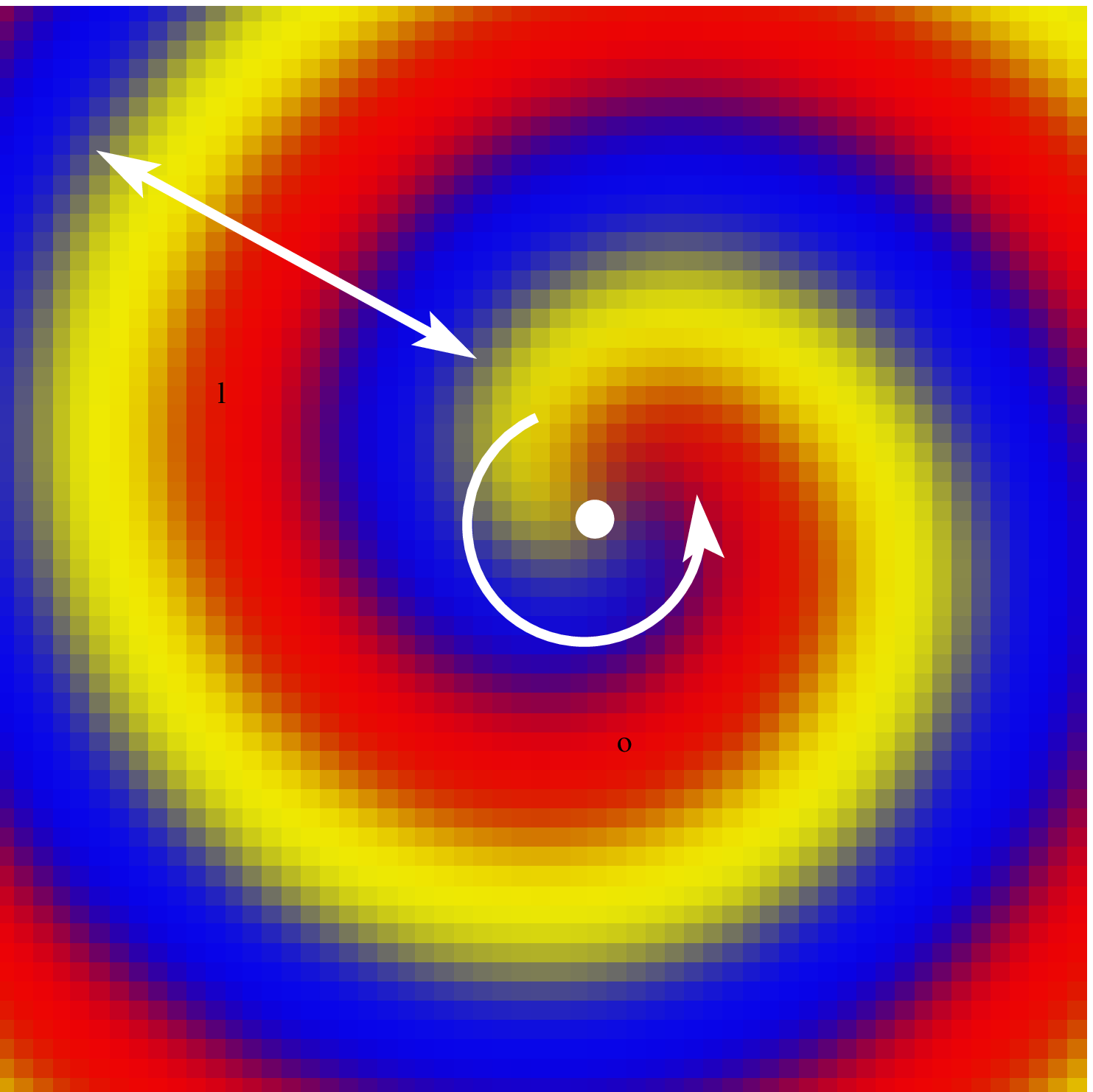}&
\includegraphics[scale=0.2]{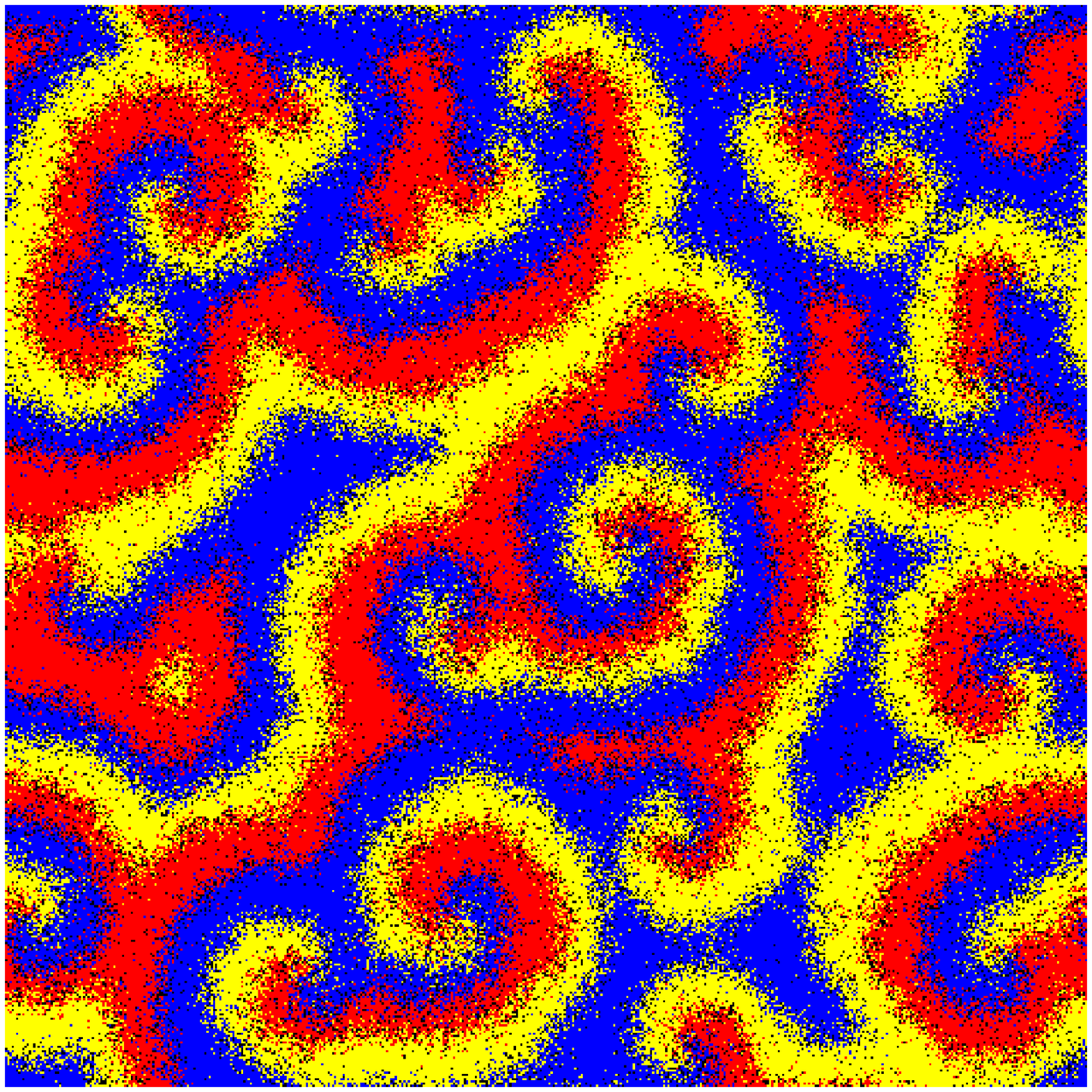}&
\includegraphics[scale=1, angle=90]{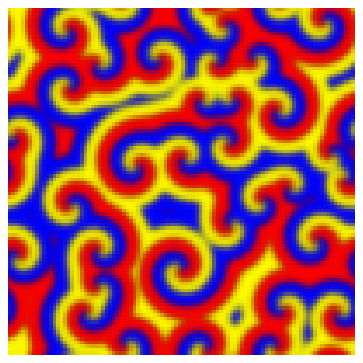}&
\includegraphics[scale=0.5]{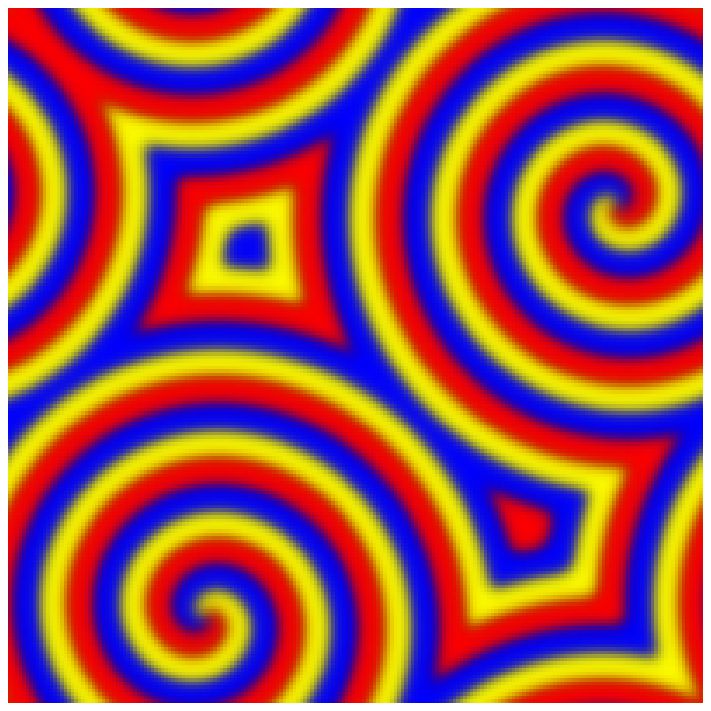}
\end{tabular}
\end{center}                
\end{figure}

\newpage

   Figure 4

\vspace{2cm}
  
\begin{figure}[h]  
\begin{center}    
\psfrag{unstable}{\sffamily Uniformity}
\psfrag{stable}{\sffamily Biodiversity}
\psfrag{-3}{\hspace{-1.1cm}\footnotesize$1\times 10^{-3}$}
\psfrag{-4}{\hspace{-1.1cm}\footnotesize$1\times 10^{-4}$}
\psfrag{-5}{\hspace{-1.1cm}\footnotesize$1\times 10^{-5}$}
\psfrag{-2}{\footnotesize$0.01$}
\psfrag{0}{\footnotesize$1$}
\psfrag{2}{\footnotesize$100$}
\psfrag{Dc}{\begin{rotate}{90}\hspace{-1.7cm}\sffamily  Critical mobility,~ $M_c$\end{rotate}}
\psfrag{betagamma}{\hspace{-2cm}\sffamily Reproduction rate,~ $\mu$}
\includegraphics[scale=0.7]{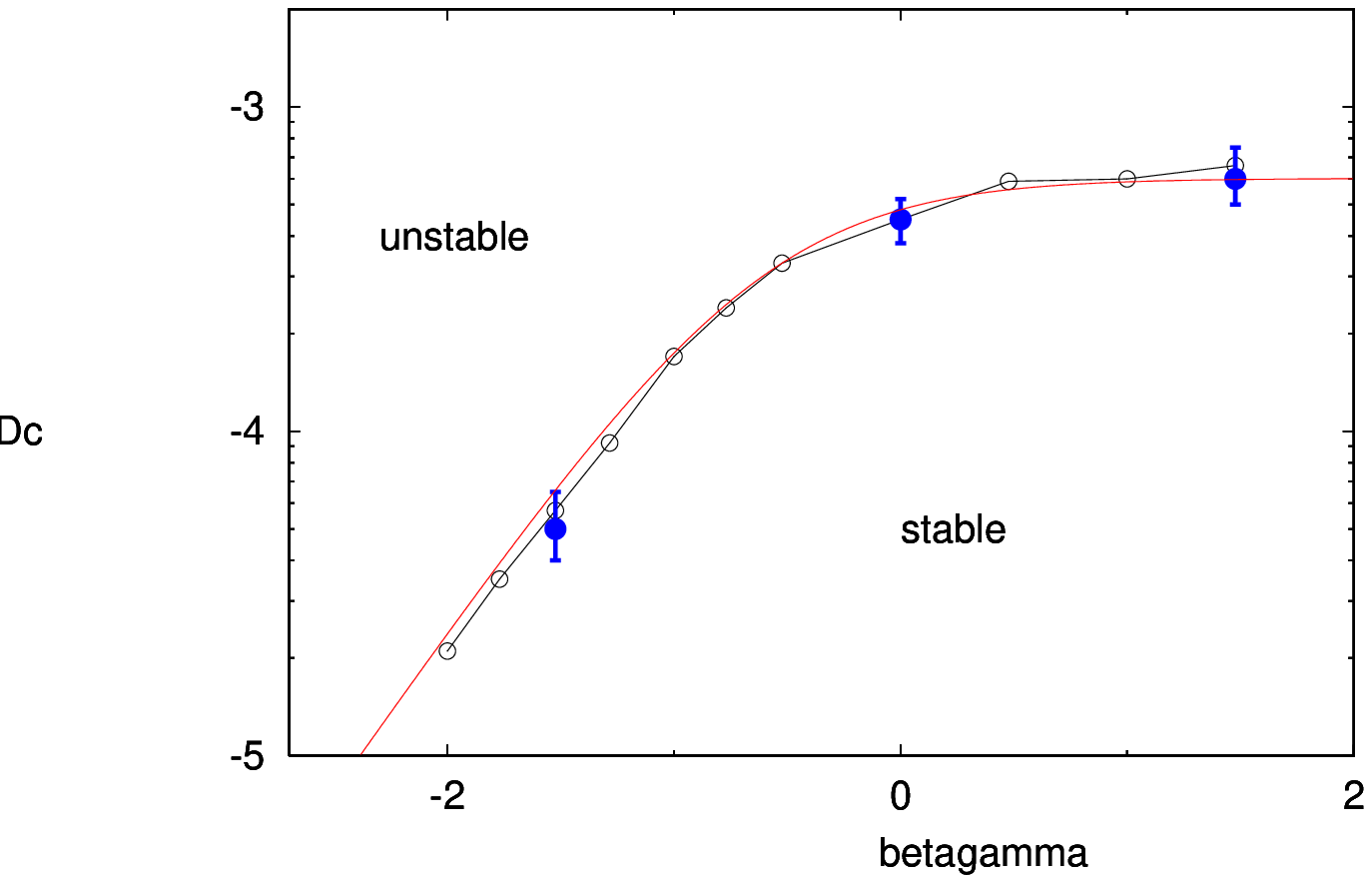}
\end{center}               
\end{figure}

\clearpage

\newpage 

%%%%%%%%%%%%%%%%%%%%%%%%%%%%%%%%%%%%%%%%%%%%%%%%%%%%%%%%%%%%%%%%%%%%%%%%%%%%%%%%%%%%%%%%
% Supplementary information
%%%%%%%%%%%%%%%%%%%%%%%%%%%%%%%%%%%%%%%%%%%%%%%%%%%%%%%%%%%%%%%%%%%%%%%%%%%%%%%%%%%%%%%%

\begin{center}
{\Large \bf 
Mobility promotes and jeopardizes biodiversity in rock-paper-scissors games
}\\
\vspace{0.5cm}
Tobias Reichenbach, Mauro Mobilia, and Erwin Frey
\end{center}

\vspace{0.5cm}

{\large \bf Supplementary Information}

\vspace{0.5cm}
In this Supplementary Information, we further elaborate our analysis by explaining some technical aspects of the Letter
and illustrate our findings by providing two supplementary movies. The latter illustrate the spatio-temporal dynamics of the biodiverse reactive states occurring in the system under consideration. \\ Below, we first present the concept of extensivity, which allows to precisely discriminate between the regime where biodiversity is stable (i.e. maintained) and the situation where it is unstable and the system
settles in one of the uniform (absorbing) states. Then, we present details on the stochastic differential equations as well as the complex Ginzburg-Landau equation  used to analyse the system.\\ We also show how the spirals' wavelength $\lambda$ 
is related to the mobility $M$. We explain that such a relation has allowed to derive the functional dependence $M_c(\mu)$ of the critical mobility on the reproduction rate. Finally, the main findings reported in the Letter are revisited in a supplementary discussion centered on the information conveyed by the movies.  \\

{\bf \large Extensivity}\\

Even if coexistence appears stable, as observed for low mobilities, 
there is a certain probability that two species go extinct due to possible large yet rare fluctuations. Indeed, the only absorbing states where no reactions (and therefore no fluctuations) occur, are the uniform configurations where only one species survives.  For this reason, these
are the only stable states in the long run.
However, the typical waiting time $T(N)$ until extinction occurs is generally very long when the system size $N$ is large. This suggests to consider  the dependence of the waiting time $T(N)$ on  $N$. Quantitatively, we discriminate between stable and unstable coexistence by using the concept of \emph{extensivity}, adapted from statistical physics.  If the ratio $T/N$ tends to infinity ($T/N\rightarrow \infty$) in the asymptotic limit of large systems ($N\rightarrow \infty$), the typical waiting time strongly prolongs  with $N$ (typically with an exponential dependence). This scenario is called \emph{super-extensive} or stable. On the other hand, if $T/N\rightarrow O(1)$ (i.e. the ratio approaches a finite non-zero value) that is referred to as the \emph{extensive} case, which has been shown to correspond to marginal (or neutral) stability$^{\ref{reichenbach-2006-74}}$. Instability of the coexistence state (towards a uniform one) is encountered when $T/N\rightarrow 0$ (\emph{sub-extensive} scenario), where the waiting time is short as compared to the system size. \\ 
These definitions of stability and instability (with neutral stability separating the two cases) in the presence of absorbing states 
are intimately related to the concept of transients. In fact, Hastings$^{\ref{hastings-2004-19}}$ 
suggested to study not only the ultimate fate of a system, but also to consider the  
behaviour at smaller (and probably ecologically more relevant) time scales. 
According to the definition introduced above, stable or neutrally stable coexistence implies coevolution of the populations  
for a very large number of generations.
This corresponds to the existence of extremely long-lived persistent transients$^{\ref{hastings-2004-19}}$ (super-persistent). 
It is also worth noticing that transients lasting for several generations can  occur even in the case of unstable 
coexistence. This typically happens when the number of individuals $N$ is large. \\
In the situation of Fig.~2, we have considered the extinction probability $P_\text{ext}$ that, starting from random initial conditions (i.e. spatially homogeneous configurations, with equal concentrations of each species), the system has reached a uniform state after a time $t$ proportional to the system size, i.e. $t\sim N$. In the asymptotic limit  $N\rightarrow\infty$, three distinct cases arise. In a first regime, the extinction probability tends to zero with the system size $N$. In Fig.~2, this occurs when $M<M_c$. This scenario corresponds to the superextensive situation (i.e. $T/N\rightarrow \infty$, with $N\rightarrow\infty$) where the coexistence of all populations is stable. As a second case, the extinction probability approaches a finite value  between $0$ and $1$, i.e. $T/N\rightarrow O(1)$, and we recover neutral stability. In Fig.~2,
such a behaviour arises exclusively at the vicinity of the critical mobility $M_c$. In a third regime, the extinction probability does reach the value $1$, which means that $T/N\rightarrow 0$. This is the subextensive scenario where the coexistence is unstable and biodiversity is lost. In Fig.~2, this happens
above the critical mobility, i.e. when $M>M_c$.\\

\newpage

{\bf \large Stochastic partial differential equations}\\
 
Within the theory of stochastic processes$^{\ref{Gardiner}}$, the dynamics of the stochastic lattice system is described by a master equation. In the limit of large systems, using a Kramers-Moyal expansion, the latter allows for the derivation of a proper Fokker-Planck equation, which in turn is equivalent to a set of stochastic partial differential equations. The latter consist of a mobility term, nonlinear terms describing the deterministic dynamics of the nonspatial model (May-Leonard equations), and noise terms. For the noise terms, we have found that contributions stemming from selection and reproduction events scale as $N^{-1/2}$, while fluctuations originating from exchanges (mobility) decay as $N^{-1}$; the latter may therefore be ignored. What remains, is (multiplicative) white noise whose strength scales as $N^{-1/2}$.\\

{\bf \large Complex Ginzburg-Landau Equation (CGLE)}\\

Ignoring the noise terms in the stochastic differential equations describing the system, the resulting partial differential  equations fall into the class of the Poincar$\acute{\text{e}}$-Andronov-Hopf bifurcation, known from the mathematics literature$^{\ref{Wiggins}}$.  Applying the theory of center manifolds and normal forms developed there, we have been able to cast the deterministic equations into the form of the complex Ginzburg-Landau equation: 
\begin{equation}
\partial_t z=M\partial_r^2 z + c_1 z -(1-ic_3)|z|^2z~,
\end{equation}
where $z$ is a complex variable and $c_1,c_3$ are constants depending on the rates $\sigma$ and $\mu$. This equation leads to the formation of dynamic spirals and allows to derive analytic results for their wavelength and frequency, see e.g. the review by Aranson and Kramer$^{\ref{aranson-2002-74}}$.\\

{\bf \large Scaling relation and critical mobility}\\

An important question is to understand what is the mechanism driving the transition from a stable coexistence to 
extinction at the critical mobility  $M_c$.  To address this issue, we first note that varying the mobility induces a scaling effect, as illustrated in Fig.~2. In fact,  increasing the mobility rate $M$ results in zooming into the system. As discussed above (see the main text and Methods), the system's dynamics is described by a set of suitable stochastic partial differential equations (SPDE) (T.R., M.M., and E.F., in preparation) whose basic properties help rationalize this scaling relation. In fact, the mobility enters the stochastic equations through a diffusive term $M\Delta$, where $\Delta$ is the Laplace operator involving second-order spatial derivatives. Such a term is left invariant when  $M$  is multiplied by a  factor $\alpha$ while the spatial coordinates are rescaled  by $\sqrt{\alpha}$. It follows from this reasoning that  varying  $M$ into $\alpha M$ translates in a magnification of the system's characteristic size by a factor $\sqrt{\alpha}$~ (say $\alpha>1$).
  This implies that the spirals' wavelength $\lambda$  is proportional to $\sqrt{M}$ (i.e. $\lambda \sim \sqrt{M}$) up to the critical $M_c$ . 

When the spirals have a critical wavelength $\lambda_c$, associated with the mobility $M_c$, these
rotating patterns outgrow the system size which results in the loss of biodiversity (see the main text). 
In the ``natural units'' (length is measured in lattice size units and the time-scale is set by keeping 
$\sigma=1$), we have numerically computed $\lambda_c=0.8\pm 0.05$. This quantity 
has been found to be universal, i.e. its value remains constant upon varying the rates $\sigma$ and $\mu$. 
However, this is not the case of the critical mobility  $M_c$, which depends on the parameters of the system. 
Below the critical threshold $M_c$, the dynamics is characterized by the formation of
spirals of wavelength $\lambda(\mu,M)\sim \sqrt{M}$. This relation, together with the universal
character of  $\lambda_c$, leads to the following 
equation:
\begin{eqnarray}
\label{Mc}
M_c(\mu)=\Big(\frac{\lambda_c}{\lambda(\mu,M)}\Big)^2M~,
\end{eqnarray}
which gives the functional dependence of the critical mobility upon the system's 
parameter. To obtain the phase diagram reported in Fig.~4 we have
used Eq.~(\ref{Mc}) together with values of $\lambda(\mu,M)$  obtained from numerical
simulations. For computational convenience, we have measured $\lambda(\mu,M)$ by carrying out a
careful analysis of the SPDE's solutions.
The results are reported as black dots in   Fig.~4. We have also  
 confirmed these results through lattice simulations  for systems with different sizes
 and the results are shown as blue dots in  Fig.~2. 
 Finally,  we have taken advantage of the analytical expression (up to a constant prefactor, taken into account in Fig.~2) of 
 $\lambda(\mu,M)$ derived  from the complex Ginzburg-Landau equation (CGLE) associated with 
 the system's dynamics (see Methods): with Eq.~(\ref{Mc}), we have obtained the red curve displayed in Fig.~2.
 This figure corroborates the validity of the various approaches (SPDE, lattice simulations and CGLE), which all lead
 to the same phase diagram where the biodiverse and the uniform phases are identified.
 
%We have therefore obtained a comprehensive phase diagram where one identifies a uniform phase, in which two species go %extinct (for $M>M_c(\mu)$), and a biodiverse phase (for $M<M_c(\mu)$) where emerging spirals induce stable coexistence of %all three species. 
\vspace{0.5cm}

{\large \bf Supplementary Movie 1}\\

In the first movie, the dynamics of individuals of species $A, B$ and $C$
follows the reactions illustrated in Fig.~1 with 
rates $\mu=1$ (reproduction), 
$\sigma=1$ (selection) and $\epsilon=2.4$ (exchange rate).
In Movie 1, individuals of each species are indicated in different colours (empty sites are shown as black dots).
The dynamics takes  place on a square lattice of $N=400\times 400$ sites, such that there are up to $1.6 \times 10^{5}$
individuals in the system. This set of parameters corresponds to a mobility rate $M=2\epsilon/N=3\times 10^{-5}$ well below the critical threshold $M_c\approx 4.5 \pm 0.5 \times 10^{-4}$ (see  Figs.~2,~4 and text).
Initially the system is in a well-mixed configuration with equal density of individuals of each species and empty sites. As time increases and since $M<M_c$, biodiversity is maintained
and complex dynamical patterns form in the course of the temporal development resulting in a rich entanglement of spiral waves.
\vspace{0.5cm}

{\large \bf Supplementary Movie 2}\\

In the second movie, the mobility of the individuals has been increased. In fact, the dynamics of individuals of species still follows the reactions illustrated in Fig.~1 with 
rates $\mu=1$ (reproduction) and 
$\sigma=1$ (selection), but the exchange rate is now $\epsilon=6$. 
In Movie 2, individuals of each species are still indicated in different colours (empty sites are shown as black dots).
 The dynamics takes 
place on a square lattice of $N=200\times 200$ sites, allowing  up to $4 \times 10^{4}$
individuals in the system. This set of parameters corresponds to a mobility rate $M=3\times 10^{-4}$ relatively close to the critical threshold $M_c\approx 4.5 \pm 0.5 \times 10^{-4}$ (see  Figs.~2,~4 and text).
Initially the system is in a well-mixed state with equal density of individuals of each species and empty sites. As time increases and since $M<M_c$, biodiversity is still maintained
and patterns form in the course of the time development. However, as compared to the first movie,
one notices that the size of the patterns has increased and one now only distinguishes one pair of antirotating spirals.

\vspace{0.5cm}

%%%%%%%%%%%%%%%%%%%%%%%%%%%%%%%%%%%%%%%%%%%%%%%%%%%%%%%%%%%%%%%%%%%%%%%%%%%%%%%%%%%%%%%%
% Supplementary discussion
%%%%%%%%%%%%%%%%%%%%%%%%%%%%%%%%%%%%%%%%%%%%%%%%%%%%%%%%%%%%%%%%%%%%%%%%%%%%%%%%%%%%%%%%

{\large \bf Supplementary Discussion}\\

The supplementary movies illustrate the system's time development in the coexistence phase, i.e.
 the emergence of dynamical complex patterns deep in that phase (Movie 1)
and close to (yet below) the threshold $M_c$ (Movie 2, see text and Fig.~3). \\
Starting from initially homogeneous
(well-mixed) configurations, after a short transient regime, spiral waves rapidly 
emerge and characterize the long-time behaviour of the system which settles in a
 reactive steady state ({\it super-extensive case}, see text). 
The wavefronts, merging to form entanglement of spirals, propagate with spreading speed
$v^{*}$ and  wavelength $\lambda$. 
In Movies 1 and 2, it appears clearly that by rising the individuals' mobility, one increases  the wavefronts propagation velocity  and
the wavelength of the resulting dynamical patterns, as well as the size of each spiral.
From PDE associated with the system's dynamics, we can rationalize this discussion and estimate these quantities
for the cases illustrated in  Movies 1 and 2. Namely, for the spreading speed, we have obtained
 $v^* \approx 3.5\times 10^{-3}$ (lattice-size units per time-step, Movie 1) and $v^*\approx 1.1\times 10^{-2}$ (Movie 2). Similarly, the wavelength were found to be $\lambda\approx
0.21$ (lattice-size units, Movie 1) and $\lambda\approx
0.65$ (Movie 2). Here, rising the rate $M$ from $3\times10^{-5}$ (Movie 1)
to $3\times 10^{-4}$ (Movie 2) results in the enhancement of $v^{*}$ and $\lambda$ by a factor
$\sqrt{10}\approx 3.16$. In Movie 2, the size of the spirals can also be  estimated
to  have been magnified by the same factor $\sqrt{10}\approx 3.16$ with respect to those of Movie 1.
As explained in the text, this scaling property of the system can be understood by considering
the stochastic partial differential equation describing the dynamics, which were obtained from the underlying
master equation through a system size expansion (see Methods).\\
By rising the individuals' mobility, one therefore increases the size of the spiralling patterns (whose wavelength is proportional to $\sqrt{M}$) and for sufficiently large value of the exchange rate (i.e. of $M$), as in Movie 2, 
just a few spirals nearly cover the entire lattice.
This happens up to the critical value $\lambda_c\approx 0.8$, found to be universal. 
In fact, when $\lambda\geq \lambda_c$
the whole system is covered with one single (``giant'') spiral which cannot fit within the lattice. This
effectively  results in the extinction of two species and the loss of biodiversity.
As explained in the text, by exploiting the fact that $\lambda\propto \sqrt{M}$ and the universal character
of $\lambda_c$, one can infer the existence of the critical mobility rate $M_c=M_c(\mu)$ [see Eq.~(\ref{Mc})], as illustrated in Fig.~4.
This allows to discuss the fate of the system (i.e. biodiversity versus extinction) in terms
of the reaction  and mobility rates $\mu$ and $M$, respectively: 
For given reaction rates $\mu$ and $\epsilon$ (without loss of generality $\sigma$ is set to unity, see text) 
and system size $L$, we obtain a critical value $M_c(\mu)$ of the mobility rate.
In fact, a reactive steady state is reached (and biodiversity maintained) only if $M<M_c(\mu)$.
When the individuals' mobility is too fast, i.e. when $M>M_c(\mu)$, the system can be considered
to be {\it well-mixed} and its dynamics therefore can be aptly described in terms of homogeneous
rate equations which predicts the extinction of two species (see Methods).\\
In the cases illustrated in Movies 1 and 2, $M_c\approx 4.5 \pm 0.5 \times 10^{-4}$ and 
the wavefronts propagate with $\lambda<\lambda_c$, so that biodiversity is always preserved.
However, we notice that the resulting spatio-temporal patterns are quite different: while one finds a rich
entanglement of spirals deep in the coexistence phase (i.e. for $M\approx 3 \times 10^{-5} \ll M_c$, Movie 1),
only one pair of antirotating spirals fill the system when one approaches the critical value $M_c$ (Movie 2).

\end{document}